\documentclass[manuscript]{aastex701}
\shorttitle{X-Ray Observations of PSR J1838--0655}
\shortauthors{Wang et al.}
\received{July 1, 2025}

\begin{document}

\title{PSR J1838--0655: X-Ray Observations with NICER and NuSTAR}

\author[orcid=0009-0001-7213-2235]{Xian-Ao Wang}
\affiliation{Department of Astronomy, Yunnan University, Kunming 650090, China}
\email{wangxianao1@stu.ynu.edu.cn}

\author[0009-0009-8477-8744]{Han-Long Peng}
\affiliation{Department of Physics and Institute of Theoretical Physics, Nanjing Normal University, Nanjing, 210023, Jiangsu, China}
\email{penghl@ihep.ac.cn}

\author{Jin-Tao Zheng}
\affiliation{Department of Astronomy, Yunnan University, Kunming 650090, China}
\email{zjt_@mail.ynu.edu.cn}

\author{Shi-Qi Zhou}
\affiliation{School of Physics and Astronomy, China West Normal University, Nanchong 637002, People’s Republic of China}
\email{Pulsar.SqZhou@gmail.com}

\author{Wen-Tao Ye}
\affiliation{State Key Laboratory of Particle Astrophysics, Institute of High Energy Physics, Chinese Academy of Sciences, Beijing 100049, China}
\email{yewt@ihep.ac.cn}

\correspondingauthor{Ming-Yu Ge}
\author{Ming-Yu Ge}
\affiliation{State Key Laboratory of Particle Astrophysics, Institute of High Energy Physics, Chinese Academy of Sciences, Beijing 100049, China}
\email{gemy@mail.ihep.ac.cn}

\correspondingauthor{Xiang-Hua Li}
\author{Xiang-Hua Li}
\affiliation{Department of Astronomy, Yunnan University, Kunming 650090, China}
\email{xhli@ynu.edu.cn}

\author{Shi-Jie Zheng}
\affiliation{State Key Laboratory of Particle Astrophysics, Institute of High Energy Physics, Chinese Academy of Sciences, Beijing 100049, China}
\email{zhengsj@ihep.ac.cn}

\begin{abstract}
We report on the timing and spectral properties of PSR J1838-0655 using joint observations from the Neutron Star Interior Composition Explorer (NICER) and the Nuclear Spectroscopic Telescope Array (NuSTAR). By disentangling the pulsar's emission from its surrounding wind nebula across joint Chandra, NuSTAR, and NICER observations, we find the pulsar's broad-band X-ray spectrum (1.3--79\,keV) is best-described by a broken power-law model. The model features photon indices of $\Gamma_1 = 1.19 \pm 0.07$ and $\Gamma_2 = 1.47 \pm 0.02$ below and above a break energy of $E_{\rm b} = 7.7 \pm 0.8$\,keV. The resulting unabsorbed 2--10\,keV flux from the pulsar is $(9.5^{+0.4}_{-0.3}) \times 10^{-12}~\mathrm{erg\,cm^{-2}\,s^{-1}}$. Furthermore, timing analysis of NICER data spanning MJD 58250 to 60630 reveals a very large glitch occurring around MJD 59300, characterized by a frequency jump of $\Delta \nu = 29.367(7) \times 10^{-6}$ Hz, which can be well explained by the vortex creep model. Phase-resolved spectral analysis indicates a clear anti-correlation between the photon index and the pulse intensity, suggesting spectral hardening at the pulse peak.

\end{abstract}



\section{Introduction} 
\label{sect:intro}
Pulsars, which are highly magnetized and rapidly rotating compact stars, serve as extraordinary laboratories for fundamental physics, offering insights into the internal structure of neutron stars, particle acceleration mechanisms, and gravitational wave emission \citep{2022hxga.book...30N}. Their rotational stability can even rival that of atomic clocks \citep{2012MNRAS.427.2780H}. Despite this remarkable stability, their long-term rotation is perturbed by two primary noise processes: continuous, long-timescale timing noise, and glitches—a sudden change in their spin frequency \citep{2008LRR....11....8L}. 
Glitches, which are sudden spin-change events, provide a unique observational window into the interior dynamics of neutron stars and the physics of superfluids under extreme conditions of density and pressure. Within the conventional neutron star, while early theories attributed glitches to crustal fractures (the starquake model; \citealp{1969Natur.223..597R}), this model struggles to explain large glitch \citep{2025ApJ...982..181Z}. The prevailing theoretical framework is the superfluid model, which posits that a glitch is caused by a sudden transfer of angular momentum from a more rapidly rotating interior neutron superfluid to the star's solid crust \citep{1975Natur.256...25A}. It is worth noting that alternative proposals for the nature of the compact objects themselves, such as solid strangeon stars, also seek to explain glitches through a starquake mechanism, but one that involves the entire stellar body rather than just a thin crust \citep{2018MNRAS.476.3303L}.
Focusing on the mainstream superfluid model, the subsequent post-glitch recovery, where the pulsar's spin frequency gradually relaxes, is a key prediction of this framework, offering deep insights into the internal coupling mechanisms. This relaxation is thought to be governed by the dynamics of quantized vortex lines within the superfluid. Specifically, the vortex creep model describes how these vortices, which are 'pinned' to nuclei in the stellar crust, unpin and move, thereby driving the recovery process \citep{1984ApJ...276..325A, 1989ApJ...346..823A}. Observationally, this process manifests as a complex relaxation, often comprising one or more exponential decay components followed by a long-term linear recovery in the spin-down rate \citep{2022Univ....8..641Z}. The detailed modeling of this recovery phase is a powerful tool for probing the microphysical properties of the inner crust, such as the fractional moment of inertia of the crustal superfluid \citep{2025arXiv250602100G}. This comprehensive framework has been particularly successful in describing the glitch cycle for prolific glitchers like the Vela pulsar (PSR J0835--4510) \citep{2020MNRAS.496.2506G, 2022MNRAS.511..425G, 2025arXiv250602100G}. Given its power in connecting microphysical conditions to observational data, we adopt the vortex creep model in this paper to analyze the post-glitch recovery.
To fully understand pulsars, multi-wavelength observations are crucial. While observations across the radio and X-ray bands are now common, the soft gamma-ray regime (tens to hundreds of keV) remains relatively unexplored. This is due to both instrumental limitations and the intrinsic faintness of most pulsars in this band \citep{2015MNRAS.449.3827K}. To date, some rotation-powered pulsars (RPPs) have been confirmed to exhibit soft gamma-ray emission.

PSR J1838-0655 is one of these rare pulsars, identified by \cite{2015MNRAS.449.3827K} as a source with soft gamma-ray emission. Its nature as a soft gamma-ray emitter makes it a compelling target for detailed investigation, promising new insights into the high-energy emission processes that are inaccessible in other pulsars. In this work, we present a comprehensive timing and spectral analysis of PSR J1838-0655, leveraging the high sensitivity and temporal resolution of current instruments to probe its behavior in detail. 

The X-ray source AX J1838.0–0655 was first detected in the Galactic plane by the Einstein satellite \citep{1988AJ.....96..233H} and was later resolved and analyzed by ASCA \citep{2001ApJS..134...77S}. Early studies associated it with the supernova remnant candidate G25.5+0.0 \citep{2003ApJ...589..253B} and identified it as a likely X-ray counterpart to the TeV source HESS J1837–069 \citep{2005Sci...307.1938A}. The pivotal discovery came from \cite{2008ApJ...681..515G}, who used RXTE data to detect a 70.5 ms pulsation. This firmly established the source as a young, energetic rotation-powered pulsar, hereafter PSR J1838–0655, with a characteristic age of 23 kyr, a surface magnetic field of $1.9 \times 10^{12}$ G, and a spin-down luminosity of $\dot{E} = 5.5 \times 10^{36}$ erg s$^{-1}$. Subsequent high-resolution Chandra imaging resolved the system into a central point source surrounded by a distinct pulsar wind nebula (PWN). 
The broad-band spectral properties of PSR J1838–0655 have been a subject of extensive study, revealing a complex picture. Early observations with ASCA and INTEGRAL were adequately described by a single power-law model, though the best-fit photon index varied significantly between observations, with reported values of $\Gamma \approx 0.5-0.8$ below 10\,keV \citep{2001ApJS..134...77S, 2003ApJ...589..253B} and $\Gamma = 1.5$ in the broad 1–300\,keV band \citep{2005ApJ...630L.157M}. A more complex picture emerged with the work of \cite{2009MNRAS.400..168L}, who analyzed combined Chandra, RXTE, and Suzaku data. They discovered a spectral break at approximately 6.5\,keV, with the photon index steepening from $\Gamma_{\rm 1} = 1.0$ to $\Gamma_{\rm 2} = 1.5$. Critically, they noted that the pulsar's 0.8–10\,keV luminosity corresponds to a spin-down conversion efficiency of ~0.9\%, unusually high compared to other pulsars. They proposed this high efficiency could be explained if PSR J1838–0655 is a highly inclined rotator. Later XMM-Newton observations, though limited by the source's off-axis position, yielded $\Gamma = 1.25$ \citep{2012ApJ...745...99K}. 

A systematic timing campaign using RXTE data from 2008 to 2010 led to the discovery of a large spin-up glitch that occurred between MJD 55002 and 55018, with a large fractional frequency change of $\Delta\nu/\nu = 1.55 \times 10^{-6}$ \citep{2010ATel.2446....1K}. The most recent comprehensive study by \cite{2015MNRAS.449.3827K} utilized combined RXTE and INTEGRAL data to present detailed, energy-resolved pulse profiles up to 150\,keV, confirming its double-peaked structure. However, since these last detailed investigations, this intriguing source has not been systematically studied using data from the current generation of high-sensitivity X-ray instruments, despite the availability of such observations. 

We have performed a comprehensive analysis of NuSTAR and NICER observations of PSR J1838–0655 to investigate the long-term evolution of PSR J1838--0655 over the past several years. This paper is organized as follows. In Section 2, we detail the observations and the data reduction procedures for our timing and spectral analyses. The results, including the discovery of a new glitch and the characterization of the pulsar's long-term spectral evolution, are presented in Section 3. Finally, in Section 4, we summarize our findings and discuss their physical implications for our understanding of this pulsar. 

\section{OBSERVATIONS AND DATA REDUCTION} \label{sect:Obs}

\subsection{Data Reduction}
\subsubsection{NICER}
\label{sect:nicer_data}

The Neutron Star Interior Composition Explorer (NICER) is an X-ray timing instrument on the International Space Station, offering high sensitivity and sub-microsecond timing resolution in the 0.2–12\,keV \citep{2016SPIE.9905E..1IP}. Between May 2018 and November 2024, NICER observed PSR J1838–0655 in 167 separate pointings, accumulating a total exposure of approximately 356 ks. We used all available data within this period. 
Data were processed using HEASoft (v6.34) \citep{2014ascl.soft08004N} and the NICER Data Analysis Software with the CALDB release xti20240206. Since NICER does not provide imaging capability, we analyzed data from the entire extraction region, including contributions from the PWN. During the data analysis, we first applied \texttt{nicerl2} to perform preliminary processing on all NICER observations, during which no issues were identified. We then used the \texttt{nicerl3-spect} command, selecting the 3C50 model for background estimation to generate spectra. While running \texttt{nicerl3-spect}, we found that 10 observations exited with error status 218. According to NICER data processing documentation, this error is due to the absence of Good Time Intervals (GTIs)\footnote{\url{https://heasarc.gsfc.nasa.gov/docs/nicer/analysis\_threads/nicerl3-spect/}}, and thus these observations were discarded.

The spectra were regrouped using \texttt{grppha} with a minimum of 70 counts per bin. After inspecting the rebinned spectra, we selected 67 observations with high signal-to-noise ratios within this energy range for spectral analysis. Each observation was individually fitted using the tbabs*power-law model at 1.3--10\,keV, and the spectral parameters were found to be generally stable over time without significant variations.
We extracted light curves using the \texttt{nicerl3-lc} command in the 12--15\,keV with a time bin size of 50 seconds. Examination of the light curves revealed that some observations exhibited abnormally high count rates at the beginning or end of GTIs. Since PSR J1838--0655 is known to be a stable X-ray pulsar, and following confirmation from NICER team members, these anomalies were attributed to instrumental effects. We therefore applied additional GTI filtering using the \texttt{maketime} command, selecting the 12--15\,keV band and retaining only intervals with a count rate below 0.2 counts s$^{-1}$.
Based on the refined GTIs, we reran \texttt{nicerl2} with enhanced filtering options (\texttt{saafilt="YES"}, \texttt{nicersaafilt="NO"}). The cleaned observations were merged using the \texttt{niobsmerge} command, and the merged dataset was processed again with \texttt{nicerl3-spect} (adopting the 3C50 background model) to produce the final spectra used in our study.

\subsubsection{NuSTAR}
\label{sect:nustar_data}

The Nuclear Spectroscopic Telescope Array (NuSTAR) is the first focusing hard X-ray telescope, composed of two co-aligned modules (FPMA and FPMB) sensitive in the 3–79\,keV \citep{2013ApJ...770..103H}. PSR J1838–0655 was observed by NuSTAR on two occasions in October 2019 (ObsIDs: 30501013002, 30501013004).
Data were reduced using the NuSTAR Data Analysis Software and CALDB v20250415. Due to NuSTAR's angular resolution (18\arcsec{} FWHM), the pulsar and its PWN could not be spatially resolved. We therefore extracted the source spectrum from a circular region of 60\arcsec{} radius centered on the pulsar's position \citep{2008ApJ...681..515G}. The background was estimated from a nearby, source-free annular region with inner and outer radii of 60\arcsec{} and 120\arcsec{}, respectively. Spectra and event files were extracted for both FPMA and FPMB using the \texttt{nuproducts} task.

\subsection{Timing Analysis}
\label{sect:TimingAnalysis}

The first step for all timing analysis was to convert photon arrival times to the Solar System Barycenter reference frame. This was performed using the \texttt{barycorr} task, which corrects for all relevant orbital and relativistic effects, referencing the times to the Barycentric Dynamical Time (TDB) standard with the JPL DE430 solar system ephemeris.

The rotational evolution of the pulsar is modeled by folding the barycenter-corrected event times according to the Taylor expansion:
\begin{equation}
\phi(t) = \phi_{\rm 0} + \nu (t - t_{\rm 0}) + \frac{1}{2} \dot{\nu} (t - t_{\rm 0})^2 + \frac{1}{3!} \ddot{\nu} (t - t_{\rm 0})^3
\label{eq:timing_model}
\end{equation}
where $\phi_{\rm 0}$ is the phase at a reference epoch $t_{\rm 0}$, and $\nu$, $\dot{\nu}$, and $\ddot{\nu}$ are the spin frequency and its first two derivatives.
We constructed a phase-coherent timing solution for the entire NICER dataset using the tempo2 software package \citep{2006MNRAS.369..655H}.
Due to the low photon statistics of some NICER observations, we first grouped the data for analysis. For the early observations (MJD 58260--59300), we merged exposures to ensure each resulting data group contained sufficient photon counts for a robust analysis (typically $>$ 2500). For observations after MJD 59300, most were treated as individual groups owing to their better photon statistics, while the few with poor statistics were merged with adjacent observations.
The pulse times of arrival (ToAs) were derived through the following procedure, similar to methods described in \citet{Tuo2020} and \citet{2019NatAs...3.1122G}:
(a) After establishing MJD 58260 as the global reference epoch and using the solution from \citet{2015MNRAS.449.3827K} as an initial guess for the timing analysis, we performed a local frequency search (Pearson's $\chi^2$ test) within each data group to determine its own optimal spin frequency. (b) Using this locally determined optimal frequency, the photon data of the group were folded to generate its pulse profile. (c) This profile was then cross-correlated with a global template of high statistical significance, and the resulting phase shift was used to calculate the ToA for the data group. This global template was initially based on the single-group profile with the best data quality, and was later regenerated by co-adding all aligned profiles after a preliminary timing solution was obtained, so as to further improve its precision. (d) The uncertainty of each ToA was estimated using a Monte Carlo method.
We repeated the procedure described in steps (a)--(d) for all data groups, thereby generating a complete set of ToAs. Finally, this entire dataset, including all ToAs and their uncertainties, was input into the tempo2 software package for an iterative fitting of the timing model to get best-fit timing solution.
Given the short time span of the two NuSTAR observations, they were not used for ephemeris fitting. Instead, their data were folded using the final NICER-derived ephemeris to generate pulse profiles. To ensure absolute consistency, we explicitly re-ran \texttt{barycorr} on the NuSTAR event files using the same ICRS reference frame and DE430 ephemeris as for the NICER data before folding.

\subsection{Phase-Resolved Spectral Analysis}
\label{sect:PhaseResolvedSpectral}

Using the final timing solution from Section~\ref{sect:TimingAnalysis}, pulse phases were assigned to each barycenter-corrected NuSTAR event. The data were then divided into 10 phase bins. For each bin, a phase-resolved spectrum was extracted using the same source and background regions defined in Section~\ref{sect:nustar_data}. These spectra were then fitted with the same model used for the phase-averaged analysis, with the interstellar absorption column density fixed to the phase-averaged best-fit value. This allowed us to investigate the variation of the spectral parameters, particularly the photon index, as a function of the pulse phase.

\section{Results}
\label{sect:result}

\subsection{Timing Result}
\subsubsection{Profile}

\begin{figure}
   \plotone{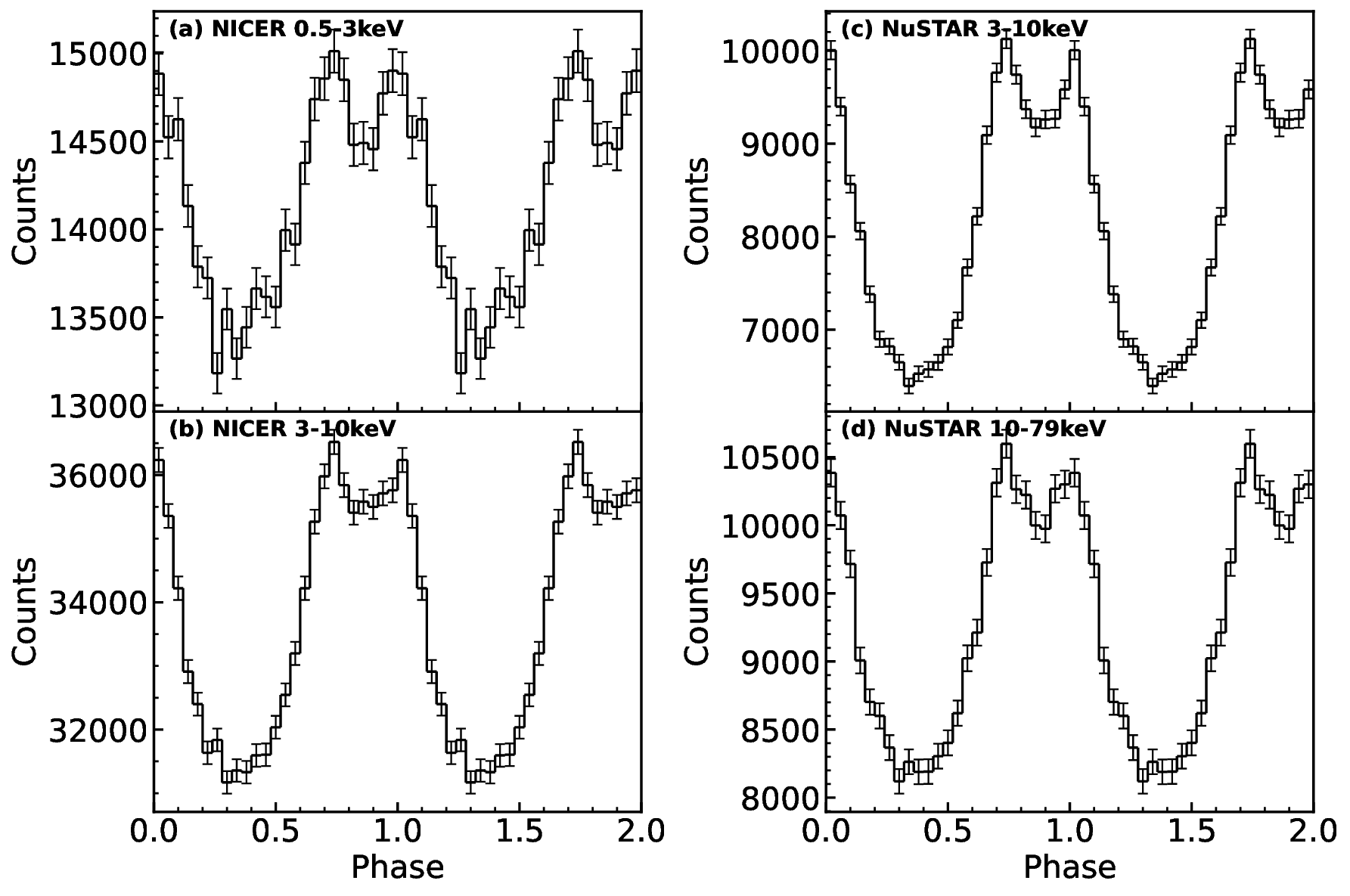}
   \caption{Pulse profiles observed by NICER and NuSTAR are presented. For NICER, the profiles were derived by stacking observations spanning MJD 58250 to 60630. Panel (a) shows the 0.5–3\,keV profile, and Panel (b) shows the 3–10\,keV profile. For NuSTAR, the profiles were obtained by combining two distinct observations and merging data from the FPMA and FPMB detectors. Panel (c) presents the 3–10\,keV profile, and Panel (d) shows the 10–79\,keV profile. All profiles are divided into 25 phase bins, with two full cycles plotted for each.
}
   \label{Fig1}
\end{figure}

We performed pulsar timing analysis by dividing NICER observations into three segments: MJD 58250--59300, MJD 59300--59600 and 59600-60630. The timing parameters for each interval were determined following the procedures described in Section~2.3. For pulse profile analysis, we selected two energy bands (0.5--3\,keV and 3--10\,keV) and generated pulse profiles for each individual observation within these intervals, which the resulting profiles, each divided into 25 phase bins.
Pulse profiles within each NICER segment were stacked according to their corresponding timing solutions to enhance the signal-to-noise ratio. Subsequently, profiles from different segments were phase-aligned and combined. The reference profile for alignment was derived from the first NICER segment (MJD 58250--59300), characterized by timing parameters at MJD~58260: $\nu = 14.181581604(1)\,\mathrm{Hz}$, $\dot{\nu} = -9.97225(8) \times 10^{-12}\,\mathrm{Hz\,s^{-1}}$, and $\ddot{\nu} = 3.71(2) \times 10^{-22}\,\mathrm{Hz\,s^{-2}}$. As no noticeable changes in the pulse profile were detected across energy or time, we present the final combined NICER pulse profiles in Figure~\ref{Fig1}.

Since two NuSTAR observations both occurred within the first NICER segment (MJD~58250--59300), we folded the NuSTAR data using the timing solution obtained for this interval. Pulse profiles were produced in two energy bands (3--10\,keV and 10--79\,keV) for both FPMA and FPMB, each divided into 25 phase bins, consistent with NICER profiles. Cross-correlation analysis and visual inspection revealed that NuSTAR pulse profiles from these two observations were well aligned. Hence, these profiles were combined to create the final NuSTAR pulse profiles, as displayed in Figure~\ref{Fig1}.

\subsubsection{Glitch}
Based on the derived ephemeris, we calculated the characteristic age of PSR J1838--0655 to be approximately 23 kyr and the surface magnetic field strength to be about $1.9 \times 10^{12}$ G, which are consistent with the results reported by \citet{2008ApJ...681..515G}. 
We identified a very large glitch in PSR J1838--0655 occurring at MJD 59300, with a frequency jump of $29.367(7) \times 10^{-6}$\,Hz, which is an order of magnitude larger than the glitch reported by \citet{2010ATel.2446....1K} in 2010.

\begin{figure}
    \plotone{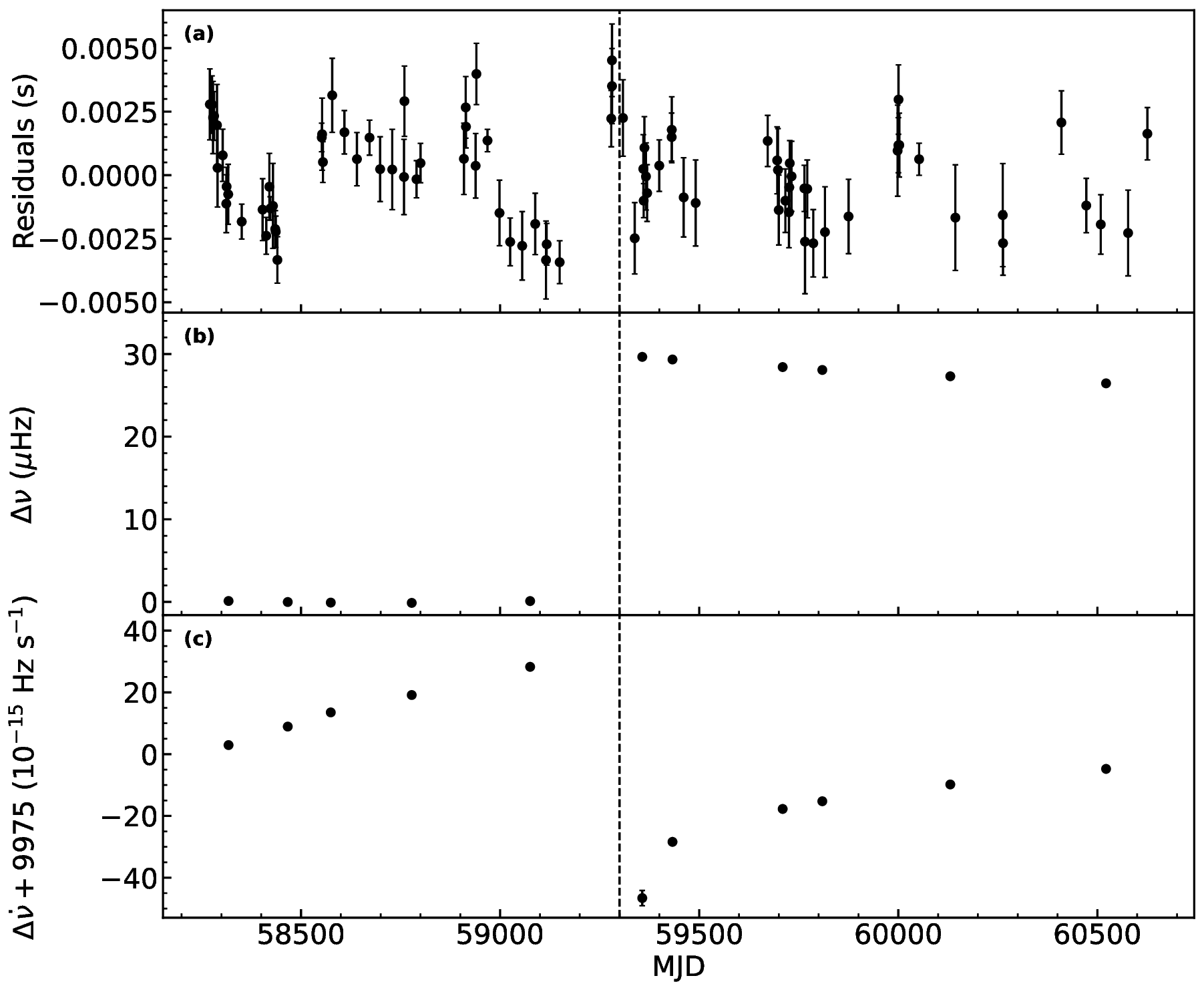}
    \caption{Step-timing analysis of PSR J1838--0655 using NICER data. Panel (a): the best fit timing residuals during MJD 58250--60630. Panel (b):frequency residuals after subtracting the pre-glitch timing model. Panel (c): the evolution of $\dot{\nu}$. The glitch epoch at MJD $59300.00(2)$ is indicated by the black dashed line.}
    \label{Fig2}
\end{figure}

To analyze the glitch, we followed the method used by \citet{2010ApJ...719L.111Y} in their study of PSR B2334+61. The pulse frequency evolution due to a glitch can be modeled using the change in frequency components, including $\Delta \nu_p$, $\Delta \dot{\nu}_p$, $\Delta \ddot{\nu}_p$, and an exponential decay term over a characteristic timescale. The total frequency $\nu(t)$ at time $t$ after the glitch is expressed as:
\begin{equation}
\nu(t) = \nu_{\rm 0}(t) + \Delta \nu_{\rm p} + \Delta \dot{\nu}_{\rm p} t + \frac{1}{2} \Delta \ddot{\nu}_{\rm p} t^2 + \Delta \nu_{\rm d} e^{-t/\tau_{\rm d}}
\end{equation}
where $\Delta \nu_{\rm p}$, $\Delta \dot{\nu}_{\rm p}$, and $\Delta \ddot{\nu}_{\rm p}$ represent the permanent changes in spin frequency and its one two derivatives at the time of the glitch. The term $\Delta \nu_{\rm d}$ corresponds to the exponential decay part, with the decay timescale given by $\tau_{\rm d}$. The total frequency change at the glitch is defined as $\Delta \nu_{\rm g} = \Delta \nu_{\rm p} + \Delta \nu_{\rm d}$. The recovery factor after the glitch is given by $Q = \Delta \nu_{\rm d} / \Delta \nu_{\rm g}$.
We first performed a stepwise timing analysis, as shown in Figure~\ref{Fig2}. From the variation in the first frequency derivative, we identified an exponential decay following the glitch. Moreover, no evidence for spin-down state transitions, as observed in some young pulsars \citep{2020ApJ...900L...7G}, was found. Based on the results of this analysis, we adopted the glitch epoch and amplitude as initial parameters and then fitted the data within one year after the glitch.
The frequency jump was initially set to $2.5 \times 10^{-5}$\,Hz and allowed to vary during the fit. Due to the presence of an exponential decay component, we initially fixed the change in the frequency derivative to the difference between the pre- and post-glitch ephemerides at MJD 59300, which is $-6.4 \times 10^{-14}$\,Hz\,s$^{-1}$. 
From the timing residuals, the decay timescale was estimated to be approximately 80 days. Given that the glitch amplitude of PSR J1838--0655 is comparable to that of PSR B2334+61, we adopted an initial decaying amplitude of $1 \times 10^{-7}$\,Hz, which yielded a reasonably good result.
We then included a change in the second frequency derivative, and allowed the frequency derivative and the decay timescale to vary freely. Through iterative fitting, we found that a single exponential decay component was sufficient to model the post-glitch residuals. The best-fit parameters are listed in Table~\ref{tab:glitch} and the best fit residuals are shown in Figure~\ref{Fig2}.
We compared the NICER spectra before and after the glitch and found that the unabsorbed flux had the similar characteristics as the crab \citep{2022ApJ...932...11Z}, there was no significant change. This lack of radiative change associated with a large glitch contrasts with the behavior observed in some other pulsars, where even small glitches can be accompanied by significant pulse profile variations, possibly linked to magnetospheric state changes triggered by the glitch event \citep{2023MNRAS.519...74Z}.

\begin{deluxetable}{ll}
\tablecaption{Glitch parameters of PSR J1838--0655 derived from NICER timing results. \label{tab:glitch}}
\tablehead{
\colhead{Parameter} & \colhead{Value}
}
\startdata
$\nu$ (Hz)                                       & 14.181581604(1) \\
$\dot{\nu}$ ($10^{-12}~\mathrm{Hz\,s^{-1}}$)      & $-9.97225(8)$ \\
$\ddot{\nu}$ ($10^{-22}~\mathrm{Hz\,s^{-2}}$)    & 3.71(2) \\
Pepoch (MJD)                                     & 58260 \\
Glitch epoch (MJD)                               & 59300 \\
Data span (MJD)                                  & 58250--60630 \\
Number of ToAs                                   & 88 \\
RMS residual ($\mu$s)                            & 1832 \\
$\Delta\nu_p$ ($10^{-6}$\,Hz)                    & 29.367(7) \\
$\Delta\nu_p/\nu$ ($10^{-6}$)                    & 2.0709(5) \\
$\Delta\dot{\nu}_p$ ($10^{-14}~\mathrm{Hz\,s^{-1}}$) & $-5.92(2)$ \\
$\Delta\dot{\nu}_p/\dot{\nu}_p$ ($10^{-3}$)         & 5.95(2) \\
$\Delta\ddot{\nu}_p$ ($10^{-22}~\mathrm{Hz\,s^{-2}}$) & $-1.86(3)$ \\
$\Delta\nu_d$ ($10^{-6}$\,Hz)                    & 0.29(1) \\
$\tau_d$ (days)                                  & 78(6) \\
$\Delta\nu_g$ ($10^{-6}$\,Hz)                    & 29.657 \\
$Q$ ($10^{-3}$)                                  & 9.7 \\
\enddata
\end{deluxetable}

\subsection{Spectral Result}
\label{sect:Phase-Resolved Spectral}
The NICER spectrum was fitted using the tbabs*power-law model. The best-fit index is $\Gamma = 1.20 \pm 0.04$, with the hydrogen column density of $N_{\rm H} = (4.29 \pm 0.01) \times 10^{22}\, \text{cm}^{-2}$, resulting in a reduced $\chi^{2}/dof = 138.83/116$. The measured 2--10\,keV flux is $(1.28 \pm 0.014) \times 10^{-11}\, \text{erg}\, \text{cm}^{-2}\, \text{s}^{-1}$, which is significantly higher than the flux measured in 2008 for the pulsar plus PWN system, $9.8 \times 10^{-12}\, \text{erg}\, \text{cm}^{-2}\, \text{s}^{-1}$. We also attempted to fit the spectrum using the broken power-law model. However, no evidence for a spectral break near $6.5$\,keV, or any significant Fe emission line was detected, as reported by \cite{2009MNRAS.400..168L}. And we note that the estimated background using the 3C50 model is comparable to---or even exceeds---the source emission below 2\,keV and above 7\,keV, indicating that the 3C50 model may not accurately reproduce the true background in this observation. We also tested the SCORPEON background model, and found that the results remain largely consistent with those obtained using the 3C50 background.

For the two NuSTAR observations, we independently fitted the 3--79\,keV spectra and found no significant differences between them. We first adopted the model constant*tbabs*power-law, with the constant parameter for FPMA fixed at 1. For ObsID 30501013002, the best-fit index was $\Gamma = 1.47 \pm 0.01$ and the hydrogen column density was $N_{\rm H} = (6.4 \pm0.4) \times 10^{22}~\mathrm{cm}^{-2}$, with a reduced chi-square of $\chi^2/\mathrm{dof} = 291.07/249$. For ObsID 30501013004, we obtained $\Gamma = 1.46 \pm 0.02$ and $N_{\rm H} = (6.2 \pm 0.5) \times 10^{22}~\mathrm{cm}^{-2}$, with $\chi^2/\mathrm{dof} = 134.37/129$. We also applied the broken power-law model to the spectra, but similarly did not detect the spectral break reported by \citet{2009MNRAS.400..168L}.

\begin{figure}
   \plotone{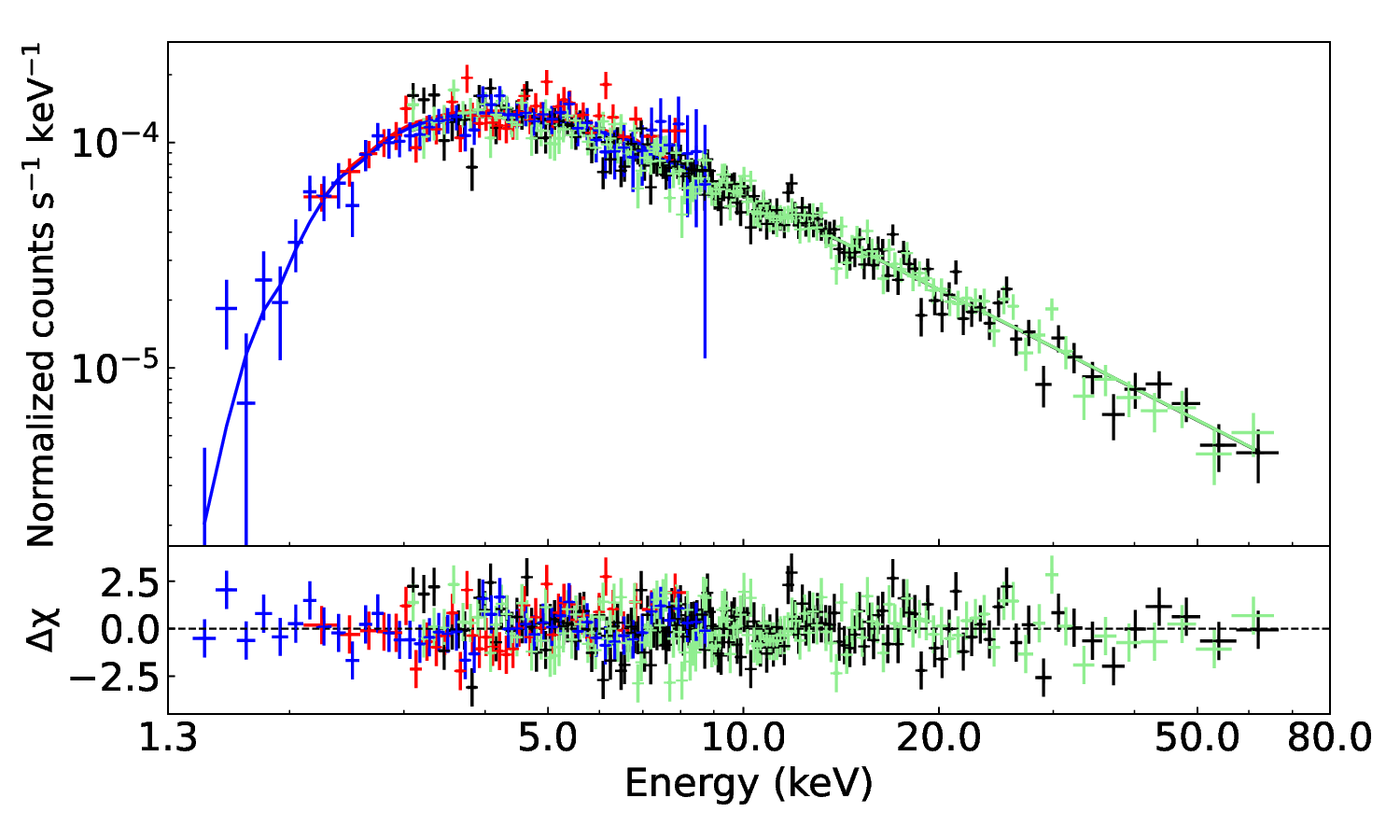}  
   \caption{The unfolded spectra observed by NuSTAR, Chandra and NICER. The top panel shows the joint spectral fit of Chandra, NuSTAR and NICER data, while the bottom panel presents the residuals. The red crosses represent the Chandra spectrum in the 2--10\,keV band, the black and green crosses correspond to the NuSTAR FPMA and FPMB spectra in the 3--79\,keV, and the blue crosses correspond to the NICER spectra in the 1.3--9\,keV, respectively. The solid lines represent the best-fit model curves for each instrument.
}
   
   \label{Fig4}
\end{figure}

The spectra from NICER and NuSTAR are unavoidably contaminated by emission from the PWN. To mitigate this contamination and extract the intrinsic spectrum of the pulsar, we processed the data as follows: (a) leveraging the high spatial resolution of Chandra~\dataset[DOI: 10.25574/cdc.431]{https://doi.org/10.25574/cdc.431}, we used the spectral data to resolve the pulsar from the PWN following the procedure of \cite{2008ApJ...681..515G} and extracted the spectrum corresponding to the pulsar; (b) for the NICER and NuSTAR data, we adopted the method used for PSR J1838-0655 in \cite{2024ApJ...965..126T} by using our resulting ephemeris to define the 0.4-phase-interval with the lowest counts as the off-pulse region (background) and the remaining 0.6-phase-interval as the on-pulse region (source) to re-extract the NICER and NuSTAR spectra. The final NuSTAR spectrum was combined from two observations, while the NICER spectra were merged using the \texttt{niobsmerge} and \texttt{nicerl3-spect} tasks. It should be noted that for the NICER data, we adopt the 1.3--9\,keV for this analysis due to signal-to-noise considerations. The resulting spectra were then fitted with both a power-law and a broken power-law model to calculate the corresponding 2--10\,keV flux. The best-fit parameters are presented in Table~\ref{tab:spectrum}. An F-test comparing the broken power-law model to the simpler power-law model yields a p-value of $4.15 \times 10^{-5}$, indicating that the addition of a spectral break provides a statistically significant improvement to the fit. The best-fit model is shown in Figure~\ref{Fig4}, and our resulting parameters are consistent with those reported by \cite{2024ApJ...965..126T}.

\begin{deluxetable*}{lcccccc}
\tablecaption{Best-fit Spectral Parameters for PSR J1838-0655 from NICER, NuSTAR (FPMA/B), and Chandra Data.\label{tab:spectrum}}
\tablehead{
\colhead{Model\tablenotemark{a}} & \colhead{$N_{\rm H}$} & \colhead{$\Gamma_1$} & \colhead{$E_{\rm b}$} & \colhead{$\Gamma_2$} & \colhead{$\chi^2$/d.o.f.} & \colhead{Flux\tablenotemark{b}} \\
\colhead{} & \colhead{($10^{22}$ cm$^{-2}$)} & \colhead{} & \colhead{(keV)} & \colhead{} & \colhead{} & \colhead{($10^{-12}$ erg cm$^{-2}$ s$^{-1}$)}
}
\startdata
Power-law         & $7.2 \pm 0.2$  & $1.42 \pm 0.02$ & \nodata         & \nodata         & 538.5/483 & $10.3 \pm 0.3$ \\
Broken power-law  & $6.2 \pm 0.3$  & $1.19 \pm 0.07$ & $7.7 \pm 0.8$   & $1.47 \pm 0.02$ & 518.3/481 & $9.5^{+0.4}_{-0.3}$ \\
\enddata
\tablenotetext{a}{The models fitted are constant*tbabs*powerlaw and constant*tbabs*broken-powerlaw, respectively.}
\tablenotetext{b}{Unabsorbed flux in the 2--10\,keV.}
\end{deluxetable*}

\subsection{Phase Spectrum}
We divided the phase-folded profile shown in Figure~\ref{Fig1} into ten phase intervals and extracted phase-resolved spectra from the two NuSTAR observations accordingly. To improve the signal-to-noise ratio, we combined the two observations using the \texttt{addspec}, \texttt{addarf}, and \texttt{addrmf} tasks. After merging, the spectra were regrouped using \texttt{grppha}, requiring a minimum of 50 counts per bin in the 3--79\,keV to generate the final phase-resolved spectra.

\begin{figure}
    \plotone{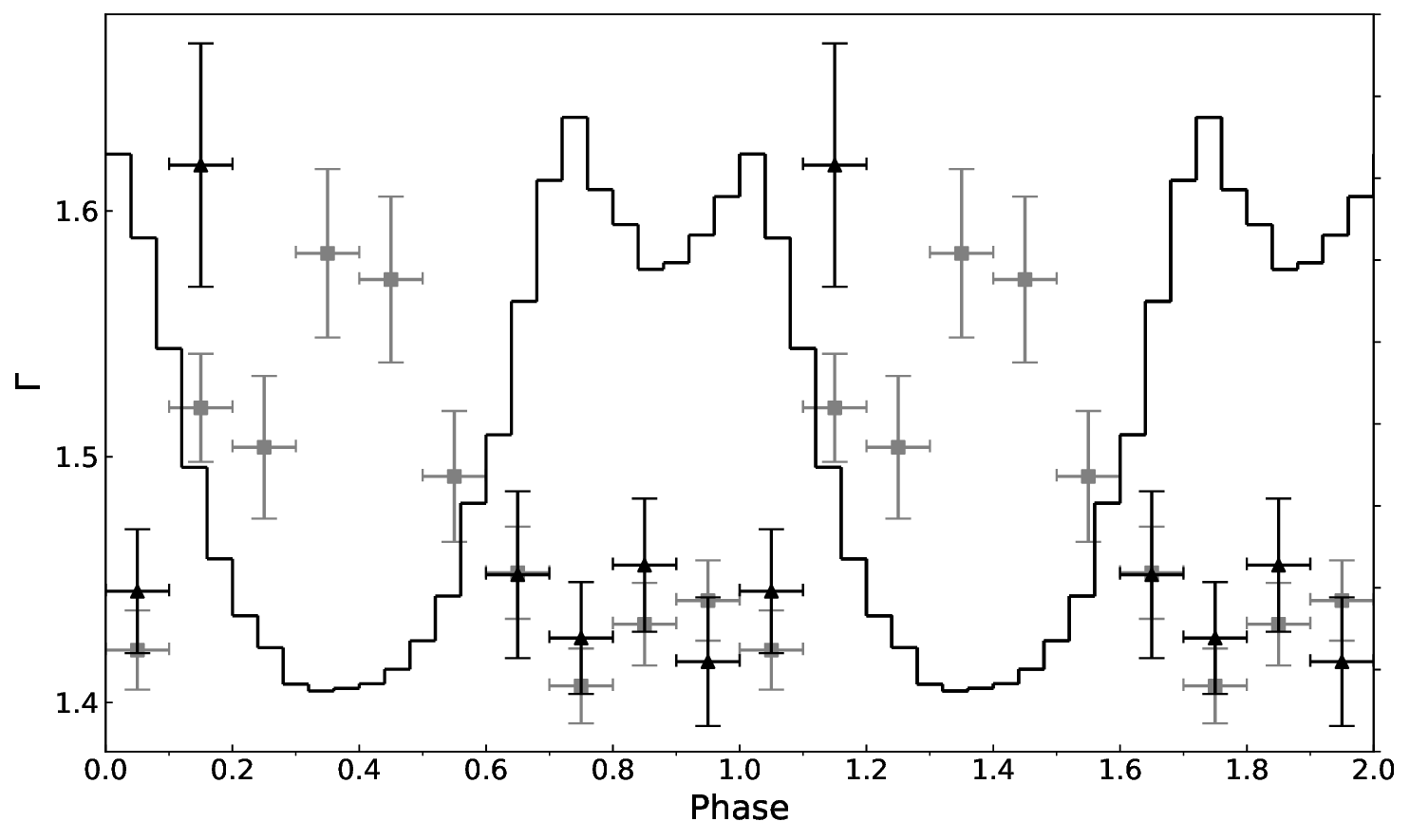}
    \caption{Phase-resolved spectral analysis of PSR J1838--0655. The black pulse profile represents the combined result of the two NuSTAR observations, stacking both FPMA and FPMB data, folded using the ephemeris of the first NICER segment in the 3--79\,keV. The profile is divided into 25 phase bins. The black triangles and gray squares represent the photon indices obtained with the background estimated from a 0.1-phase-wide off-pulse region and a 60\arcsec--120\arcsec{} annulus in 10 phase bins, respectively.}
    \label{Fig5}
\end{figure}

Spectral fitting was performed using the model constant*tbabs*power-law, with the constant for FPMA fixed at 1, and the hydrogen column density fixed at $N_{\rm H} = 6.2 \times 10^{22}~\mathrm{cm}^{-2}$. The resulting photon index as a function of phase is shown in Figure~\ref{Fig5}. A clear phase dependence is observed: the photon index is lower in the pulse phases, reaching a minimum at the main peak, then increases between the two peaks, and decreases again near the secondary peak. In the off-pulse phases, the photon index is systematically higher, peaking where the count rate is lowest. Overall, the photon index shows an anti-correlation with the pulse profile. The characteristics of this phase spectrum are similar to those of PSR B1509–58 and PSR B0540–69, which also have a hard peak and soft wings \citep{2012ApJS..199...32G}. However, this method may still be affected by residual contamination from the PWN. To test this potential systematic effect, we selected a more conservative background region, the 0.1-phase-interval off-pulse with the minimum counts, and re-derived the evolution of the spectral index using the same procedure. The resulting trend, presented in Figure~\ref{Fig5}, is consistent with our initial findings.

\section{Discussion and Conclusions}
\label{sect:discussion}

In the timing analysis of PSR J1838--0655 from MJD 58250 to MJD 60630, we observed a very large glitch at MJD 59300 and due to the limitations of the observational data, we cannot determine whether there was the delayed spin-up process \citep{2020ApJ...896...55G}. 
For the analysis of this glitch, we employ the vortex creep model to fit the post-glitch data. It is worth noting that in pulsars with more frequent and complex glitching behavior, additional phenomena, such as damped sinusoidal oscillations in the spin-down rate \citep{2024ApJ...977..243Z}, can also be observed. These are often interpreted using complementary frameworks, such as the vortex bending model.
Refer to \citet{2025arXiv250602100G} the analysis of Vela pulsar, the post-glitch spin-down rate reflects the response of the vortex creep region in both linear and non-linear states to the glitch. Under the pulsar crust, part of the superfluid exhibits a linear response to the glitch. The behavior of the spin-down rate after the glitch can be expressed by the following equation \citep{2017MNRAS.469.2313G}:
\begin{equation}
\Delta \dot{\nu}(t) = \sum_{i=1}^{3} - \frac{I_{\rm ei}}{I} \frac{\Delta \nu}{\tau_{\rm ei}} \exp \left( -\frac{t}{\tau_{\rm ei}} \right)
\end{equation}
where $\Delta \nu$ is the frequency change, $\tau_{\rm ei}$ is the timescale for the recovery of the $i^{\rm th}$ layer of the vortex, which describes the speed at which this layer returns to a stable state, $I_{\rm ei}$ is the moment of inertia of the vortex participating in the $i^{\rm th}$ layer, $I$ is the total moment of inertia of the pulsar, and $t$ is the time after the glitch. This linear response part is closely related to the exponential decay of the spin-down rate following the glitch. 

Under the pulsar crust, part of the superfluid exhibits a non-linear response to the glitch. The behavior of the spin-down rate after the glitch can be expressed by the following equation \citep{1996ApJ...459..706A}:
\begin{equation}
\Delta\dot{\nu}(t) = \frac{I_{\rm a}}{I} \nu_{\rm 0} \left[ 1 - \frac{1 - \left( \frac{\tau_{\rm nl}}{t_{\rm 0}} \right) \ln\left( 1 + \left( \frac{t_{\rm 0}}{\tau_{\rm nl}} - 1 \right) e^{-t/\tau_{\rm nl}} \right)}{1 - e^{-t/\tau_{\rm nl}}} \right]
\end{equation}
$I_{\rm a}$ is the moment of inertia of the crustal superfluid in the nonlinear creep region, $I$ is the total moment of inertia of the neutron star, $ \nu_{\rm 0} $ is the pre-glitch spin-down rate, $\tau_{\rm nl}$ is the non-linear creep relaxation time, which characterizes the timescale of the non-linear recovery, $ t_{\rm 0} $ is the offset time.
This component determines the magnitude of the glitch \citep{2022Univ....8..641Z} and is also related to the long-term recovery following the exponential decay phase.

To model the post-glitch evolution of the spin-down rate, we adopt Equations (3) and (4). 
Due to data limitations, there are only 6 step-by-step timing results after the glitch. It is not very reasonable to directly fit so many parameters. Therefore, we use the relationship between the vortex creep model parameters after exponential decay and the observable quantities of the pulsar after the transition proposed by \citet{2006MNRAS.372..489A}:
\begin{equation}
\frac{\Delta \Omega_{\rm c}}{\Omega_{\rm c}} = \left( \frac{I_{\rm a}}{2I} + \frac{I_{\rm b}}{I} \right) \frac{\delta \Omega_{\rm s}}{\Omega_{\rm c}}
\label{eq:delta_omega}
\end{equation}

\begin{equation}
\frac{\Delta \dot{\Omega}_{\rm c}}{\dot{\Omega}_{\rm c}} = \frac{I_{\rm a}}{I}
\label{eq:delta_omega_dot}
\end{equation}

\begin{equation}
\ddot{\Omega}_c = \frac{I_{\rm a}}{I} \frac{\dot{\Omega}_c^2}{\delta \Omega_{\rm s}}
\label{eq:omega_double_dot}
\end{equation}

$\Omega_{\rm c}$: Angular velocity of the pulsar's crust, $\Delta \Omega_{\rm c}$: Change in crustal angular velocity due to the glitch, $\dot{\Omega}_{\rm c}$: Post-glitch first time derivative of the crust's angular velocity, $\Delta \dot{\Omega}_{\rm c}$: Change in the angular velocity caused by the glitch, $\ddot{\Omega}_{\rm c}$: Second derivative of the angular velocity after the glitch, $\delta \Omega_{\rm s}$: Decrease in the angular velocity of the superfluid in nonlinear creep regions, $I$: Total moment of inertia of the neutron star, $I_{\rm a}$: Moment of inertia of the superfluid regions undergoing nonlinear vortex creep, $I_{\rm b}$: Moment of inertia of regions that do not participate in creep but transfer angular momentum during the glitch.

\begin{figure}
    \plotone{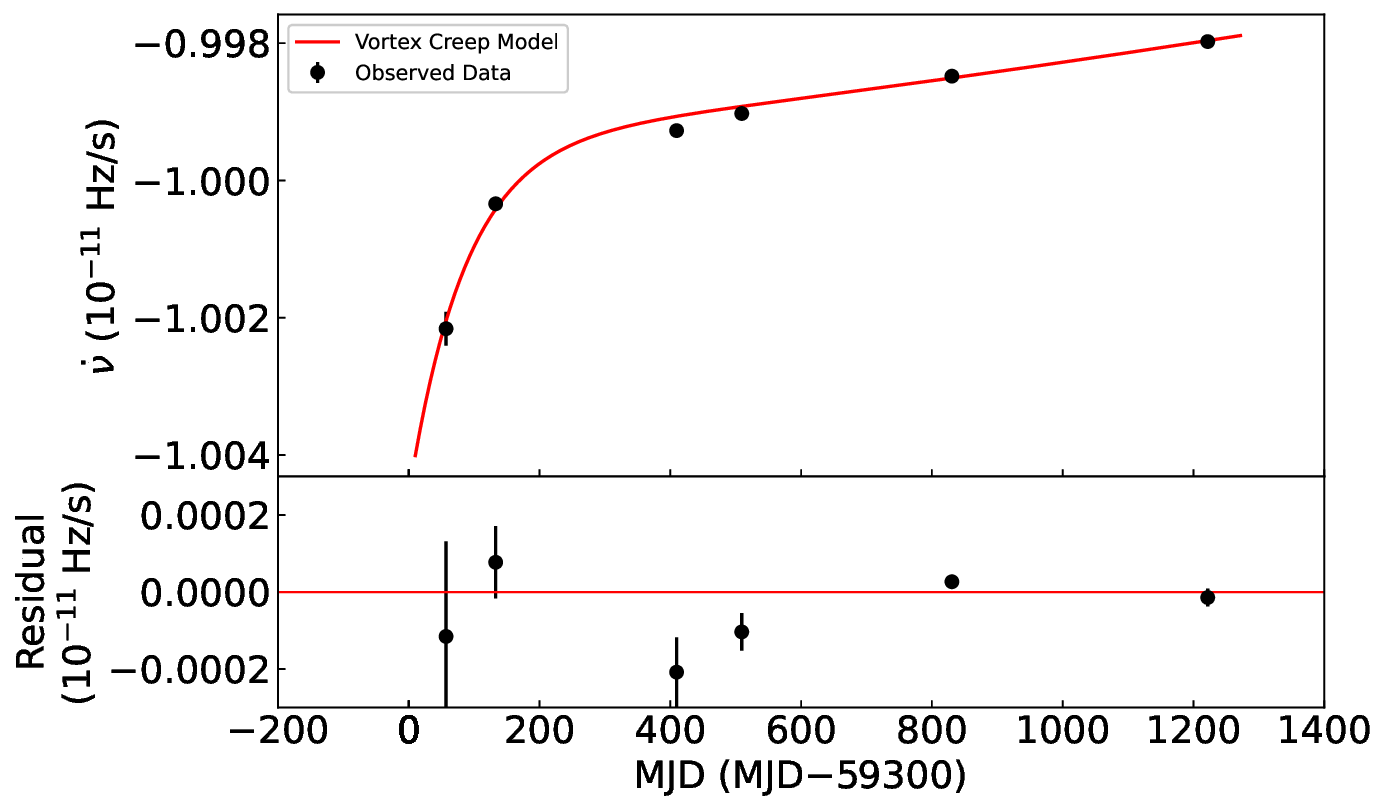}
    \caption{Horizontal axis is defined relative to the glitch epoch (MJD 59300), with zero corresponding to the glitch time. Shows the observed changes in the spin-down rate relative to the pre-glitch value (black points) and the predicted evolution from the best-fit vortex creep model (solid red line).
}
    \label{Fig6}
\end{figure}

These equations allow us to compute the fractional moment of inertia of the non-linear creep region, $I_{\rm a}/I$. From the tempo2 timing solution, we find that the post-glitch recovery can be well described using a single exponential decay component. Therefore, only one term ($i=1$) is included in Equation (3) for fitting purposes.
We first perform a least-squares fit to obtain initial estimates of the model parameters, including $I_{\rm e}/I$, $\tau_{\rm e}$, $\tau_\mathrm{nl}$, and $t_{\rm 0}$. These values provide a reasonable starting point for the MCMC simulation, which is then used to refine the parameter estimates. The final results are reported as the maximum a posteriori values along with their 95\% confidence intervals.
The final parameter values are: $I_{\rm a}/I$ = 0.00592, $I_{\rm e}/I$ =$0.013^{+0.003}_{-0.002}$ , $\tau_{\rm e} = 87^{+102}_{-19}\,\mathrm{days}$, $\tau_\mathrm{nl} = 496^{+152}_{-67}\,\mathrm{days}$, and $t_{\rm 0} = 3345^{+297}_{-133}\,\mathrm{days}$. The fitting results are shown in Figure~\ref{Fig6}. 
The vortex creep model provides a good physical explanation for the post-glitch behavior of PSR J1838-0655. The model not only provides a good fit to the overall evolution of the spin-down rate after the glitch, but its derived exponential decay timescale is also in good agreement with the result from the tempo2 fit. These results strongly support that this glitch was caused by a sudden transfer of angular momentum from the interior superfluid to the crust of the pulsar.

Joint analysis of NuSTAR, Chandra and NICER data shows that PSR J1838--0655 has an unabsorbed 2--10\,keV flux of $9.5^{+0.4}_{-0.3} \times 10^{-12}~\text{erg\,cm}^{-2}\,\text{s}^{-1}$. Assuming a distance of 6.6\,kpc, this corresponds to an X-ray luminosity of $(5.0 \pm 0.2) \times 10^{34}~\text{erg\,s}^{-1}$. Based on the timing solution prior to the glitch, the spin-down luminosity is estimated to be $L_{\rm sd} = 5.6 \times 10^{36}~\text{erg\,s}^{-1}$, consistent with the value reported by \citet{2015MNRAS.449.3827K}. The resulting X-ray conversion efficiency in the 2--10\,keV band is therefore approximately $0.89 \pm 0.03\%$, and is notably high among soft gamma-ray pulsars. This value is consistent with the result reported by \citet{2009MNRAS.400..168L}.
For comparison, PSR J1811--1925, which has a comparable characteristic age, shows an efficiency of only 0.3\% in the 1.0-10.0\,keV \citep{2015MNRAS.449.3827K}\citep{2023RAA....23k5007Z}.
In the study conducted by \citet{2004ApJ...604..317Z} \citet{2006A&A...454..537Z}, the outer gap model was employed to explain the non-thermal pulsed X-ray emission from pulsars and \citet{2009MNRAS.400..168L} further applied the outer gap model to explain the high X-ray conversion efficiency observed in PSR J1838--0655. Their analysis demonstrated that a large magnetic inclination angle significantly enhances the X-ray flux. Within the framework of this model, we find that the high conversion efficiency can be accommodated by a highly inclined rotator with a magnetic inclination angle of approximately 80 degrees or greater.

\begin{acknowledgements}
This work was financially supported by the National Natural Science Foundation of China (NSFC, Grant Nos. 12233006, 12373046, 11080922, 12373051, 12333007), the Yunnan Provincial Foundation (202301AS070073), the National SKA Program of China (2022SKA0120101), and the International Partnership Program of the Chinese Academy of Sciences (No. 113111KYSB20190020). S.Q.Z. is supported by the Key Project of the Sichuan Science and Technology Education Joint Fund (25LHJJ0097). We gratefully acknowledge the use of public data from the NICER, NuSTAR, and Chandra archives.
\end{acknowledgements}

\begin{contribution}

X.A. Wang conducted the data reduction and analysis for both NICER and NuSTAR, including spectral and timing studies, and wrote the manuscript. 
H.L. Peng, W.T. Ye, and S.Q. Zhou provided guidance on the spectral and timing analysis of NICER and NuSTAR data and assisted in resolving related issues. 
J.T. Zheng processed the Chandra data and provided instruction on fundamental X-ray data analysis techniques. 
X.H. Li, M.Y. Ge, and S.J. Zheng conceived and supervised the research, provided funding support, offered critical insights during the analysis process, and provided guidance throughout the project.


\end{contribution}

%
\facilities{NICER (XTI), NuSTAR (FPMA and FPMB), Chandra (ACIS and HRC)}

\software{ HEASoft (v6.34) \citep{2014ascl.soft08004N}; tempo2 \citep{2006MNRAS.369..655H}}


\appendix

\section{Appendix information}


\bibliography{paper}{}
\bibliographystyle{aasjournalv7}



\end{document}